\documentclass[preprint,aps,amsmath]{revtex4}

\usepackage{graphicx}
\usepackage{rotating}
\usepackage{amsfonts}

\newcommand{\du}{\Delta\hspace{-1pt} U}

\newcommand{\bal}{\begin{align}}
\newcommand{\eal}{\end{align}}

\newcommand{\eja}{E_\textrm{J1}}
\newcommand{\ejb}{E_\textrm{J2}}
\newcommand{\ejc}{E_\textrm{J3}}
\newcommand{\ecc}{E_\textrm{C3}}
\newcommand{\be}{\begin{equation}} \newcommand{\ee}{\end{equation}}
\newcommand{\ba}{\begin{eqnarray}} \newcommand{\ea}{\end{eqnarray}}
\newcommand{\bastar}{\begin{eqnarray*}}
\newcommand{\eastar}{\end{eqnarray*}} 
\newcommand{\bra}[1]{\langle #1|\,} \newcommand{\ket}[1]{|\, #1 \rangle}
  
\newcommand{\TKK}{Laboratory of Physics, Helsinki University of Technology \\ P.~O.~Box 4100, FI-02015 TKK, Finland}
\newcommand{\LTL}{Low Temperature Laboratory, Helsinki University of Technology \\ P.~O.~Box 3500, FI-02015 TKK, Finland}

\newcommand{\SAKSA}{LPMMC, CNRS, and Universit\'e Joseph Fourier, B\^oite Postale 166, 38042 Grenoble CEDEX 9, France}
\newcommand{\PISA}{NEST-CNR-INFM \& Dipartimento di Fisica, Universit\'a di Pisa, largo E. Fermi, I-56100 Pisa, Italy}
\begin{document}

\title{Measurement scheme of the Berry phase in superconducting circuits}

\author{Mikko~M\"ott\"onen}\email{mikko.mottonen@tkk.fi}\affiliation{\LTL}\affiliation{\TKK}
\author{Jukka~P.~Pekola}\affiliation{\LTL}
\author{Juha~J.~Vartiainen}\affiliation{\LTL}
\author{Valentina Brosco}\affiliation{\PISA}
\author{Frank~W.~J.~Hekking}\affiliation{\SAKSA}
\date{\today}

\begin{abstract}
We present a measurement scheme for observing the Berry phase in a flux assisted Cooper pair pump---the Cooper pair sluice. In contrast to the
recent experiments, in which the sluice was employed to generate accurate current through a resistance, we consider a device in a
superconducting loop. This arrangement introduces a connection between the pumped current and the Berry phase accumulated during the adiabatic
pumping cycles. From the adiabaticity criterion, we derive equations for the maximum pumped current and optimize the sluice accordingly. These
results apply also to the high accuracy pumping which results in a potential candidate for a metrological current standard. For measuring the
pumped current, an additional Josephson junction is installed into the superconducting loop. We show in detail that the switching of this system
from superconducting state into normal state as a consequence of an external current pulse through it may be employed to probe the pumped
current. The experimental realization of our scheme would be the first observation of the Berry phase in superconducting circuits.

\end{abstract}

\hspace{5mm} 

\maketitle
\section{Introduction}\label{sec1}
Adiabatic cyclic temporal evolution of any quantum system gives rise to geometric phases depending only on which path the system follows, not
the speed of the cycle. For quantum states in degenerate subspaces, these geometric phases correspond to non-Abelian unitary operators within
the subspace, also referred to as holonomies~\cite{Simon1983,Wilczek1984}. In holonomic quantum computation~\cite{zanardi1999}, these operators
are of special interest since they represent quantum gates necessary for the actual computation. In the non-degenerate case, the geometric phase
arising from a closed cycle in the adiabatic temporal evolution is referred to as Berry phase~\cite{Berry1984} corresponding to a phase of the
quantum state under adiabatic temporal evolution. In general, the Berry phase is not directly observable and it is observed as a phase
difference between two states which have traveled a different path during the adiabatic evolution, as was proposed in
Refs.~\cite{Falci2000,peng2006} for the yet unrealized observation of the Berry phase superconducting quantum interference devices (SQUIDs). To
date, the Berry phase has been observed~\cite{Anandan1997}, for example, in systems of electrons~\cite{Tonomura1986} circulating about a wire
carrying electric current known as Aharonov-Bohm effect, and in
systems of neutrons~\cite{Cimmino1989} or molecules~\cite{Sangster1993} circulating about a line of electric charge known as Aharonov-Casher effect. 

The peculiarity of the Berry phase in a phase biased array of Josephson junctions is that the geometric phase accumulated in the ground state of
the Hamiltonian is closely related to the charge pumped through the device. This relation was first found by Aunola et al.~\cite{Aunola2003} and
also studied in Ref.~\cite{NiskanenThesis}. Hence the yet unrealized observation of the Berry phase in superconducting circuits reduces in this
structure into detection of weak currents.

We consider a flux assisted Cooper pair pump, also referred to as Cooper pair sluice~\cite{Niskanen2003}, consisting of a superconducting island
separated by two SQUIDs which work as tunnel junctions with tunable Josephson energies. In order
to achieve large and accurate pumped currents for a novel current standard, a voltage biased sluice was employed to pump current in
Ref.~\cite{Niskanen2005}. In contrast, we place the sluice into a superconducting loop, in which the device is phase biased, and hence may be
employed to observe the Berry phase.

For the measurement of the pumped current, we suggest that an additional Josephson junction is installed in the loop in parallel with the
sluice. The pumped current modifies the switching statistics of the whole superconducting circuit into the normal conducting state under
external current pulses. This change is shown to suffice for quantitative measurements of the pumped current. Furthermore, our measurement
scenario resembles that of the superconducting qubit quantronium~\cite{Cottet2002,Vion2002}, and hence we expect it to be experimentally
feasible.

Recently, a more sensitive measurement scheme than the one implemented in Saclay~\cite{Cottet2002,Vion2002} for quantronium was implemented in
Ref.~\cite{Siddiqi2005}. In this set up, the excitation and de-excitation of the additional tunnel junction introduces phase shifts to external
current pulses, and hence no energy is transferred to the measured system. Whereas this method introduces improvements in the visibility and
contrast of the measurement, its experimental implementation is more demanding than that of the Saclay measurement
scheme~\cite{Cottet2002,Vion2002}. Moreover, we will show that the Saclay measurement scheme is actually more feasible in the observation of the
Berry phase than in the quantronium measurement, and hence well justified.

In the adiabatic temporal evolution of a quantum state, the parameters of the system Hamiltonian are controlled externally. The adiabaticity
criterion states that the change in the parameters must be slow enough for the system to stay in the same instantaneous eigenstate of the
Hamiltonian. This requirement limits the speed of pumping in the Cooper pair sluice, and hence the maximum pumped current. Thus the architecture
of the sluice must be optimized in accordance with the limitations due to adiabaticity to obtain as strong a current signal  as possible in the
proposed measurement scheme. In Ref.~\cite{Niskanen2003}, the errors in the pumped current due to the breaking of the adiabaticity were studied
using a dynamic approach for a fixed design of the sluice. In contrast, we present both numerical and analytical results for the maximum pumped
current with any error rate. The results are obtained directly from the adiabaticity theorem and the spectral analysis of the system
Hamiltonian. In addition, we apply the results to optimize the architecture of the sluice for both the Berry phase measurement and for high
accuracy current pumping.

The structure of this paper is the following: In Sec.~\ref{sec2}, we present the Hamiltonian and the circuit of the Cooper pair sluice.
Section~\ref{sec3} is devoted to the study of the Berry phase and its connection to the pumped charge. Section~\ref{sec4} presents the
optimization methods for the design of the sluice with respect to the adiabaticity criterion. The scheme to measure the pumped current is
considered in Sec.~\ref{sec5}. Section~\ref{sec6} summarizes and concludes the results of this paper.

\section{Cooper pair sluice}\label{sec2}
\begin{figure}[tbh]
\includegraphics[width=150pt]{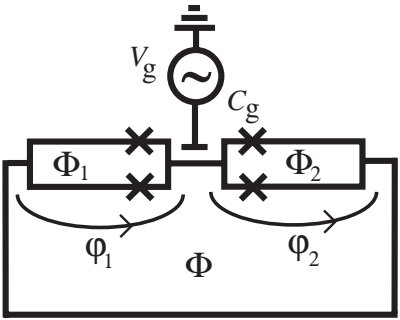}
\caption{\label{fig1} Circuit for the Cooper pair sluice consisting of a superconducting island with gate capacitance~$C_\textrm{g}$, and two
SQUIDs with magnetic fluxes~$\Phi_1$ and~$\Phi_2$ and phase differences of the order parameter~$\varphi_1$ and~$\varphi_2$. The gate voltage is
denoted by~$V_\textrm{g}$.}
\end{figure}

The Cooper pair sluice consists of a superconducting island separated by two SQUIDs. Figure~\ref{fig1} shows the sluice in a superconducting
loop. In the case of vanishing loop inductance, the SQUIDs are equivalent to Josephson junctions with tunable Josephson energies~$\eja(\Phi_1)$
and~$\ejb(\Phi_2)$, where~$\Phi_1$ and~$\Phi_2$ denote the externally controllable magnetic fluxes through the first and second SQUID,
respectively. On the other hand, the Coulomb energy for one excess Cooper pair to reside on the island is $E_\textrm{€C}=2e^2/C_\Sigma$, where
$C_\Sigma$ represents the total capacitance of the island. If one neglects the parasitic capacitances, the total capacitance is given by
$C_\Sigma=2C_\textrm{J}+C_\textrm{g}$, where~$C_\textrm{J}$ is the capacitance of one of the identical SQUIDs and $C_\textrm{g}$ is the gate
capacitance which is used to charge the island with the gate voltage $V_\textrm{g}$. Thus the device can be considered to be a tunable Cooper
pair transistor.

The Hamiltonian of the sluice may be written in the eigenbasis of the phase operator~$\phi$ as~\cite{Niskanen2003} \be\label{ham}
\hat{H}_\textrm{sl}=E_\textrm{C}(\hat{n}-n_\textrm{g})^2-\eja(\Phi_1)\cos(\phi+\varphi/2)-\ejb(\Phi_2)\cos(\phi-\varphi/2),\ee where
$\hat{n}=-i\partial_\phi$ is the number operator for the Cooper pairs on the island, $n_\textrm{g}=V_\textrm{g}C_\textrm{g}/(2e)$ is the gate
charge in units of~$2e$, $\varphi=\varphi_1+\varphi_2$ is the phase difference over the device, and the phase on the island is given by
$\phi=(\varphi_1-\varphi_2)/2$. The phase operators~$\phi$ and~$\varphi_i$ are related to the phase of the order parameter describing the
many-body quantum state of the superconductor and should not be mixed with the Berry phase or the phase of the ground state of the above
Hamiltonian. Since the sluice forms a superconducting loop, the phase~$\varphi$ over the device is fixed by the magnetic flux~$\Phi$ through the
loop as \be\label{totphase} \varphi=\frac{2\pi\Phi}{\Phi_0},\ee where $\Phi_0=\pi\hbar/e$ is the flux quantum and we have assumed the the loop
inductance is negligible. Thus~$\varphi$ may be treated as a real number.

The current operator of the $k$th SQUID is defined as \be\label{ik}
\hat{I}_k=\frac{2ie}{\hbar}[\hat{n}_k,\hat{H}]=\frac{2e}{\hbar}\frac{\partial \hat{H}}{\partial{\varphi_k}}.\ee where
$\hat{n}_k=-i\partial_{\varphi_k}$ is the Cooper pair number operator of the $k$th SQUID. For the sluice Hamiltonian Eq.~\eqref{ik} reduces into
\be\label{iks}\hat{I}_k=\frac{2eE_{\textrm{J}k}}{\hbar}\sin(\varphi_k).\ee Equation~\eqref{ik} ensures that for any states~$\ket{\xi(t)}$
and~$\ket{\chi(t)}$ of the system the current operator possesses the property
$\partial_t\bra{\xi(t)}(-2e)\hat{n}_k\ket{\chi(t)}=\bra{\xi(t)}\hat{I}_k\ket{\chi(t)}$, i.e., the matrix elements of the current operator are
given by the time derivative of the matrix elements of the charge operator. The critical current~$I_{\textrm{c}k}$ is defined as the maximum
current which can flow through the $k$th SQUID in the superconducting state. Hence, Eq.~\eqref{iks} yields a relation \be
I_{\textrm{c}k}=\frac{2\pi E_{\textrm{J}k}}{\Phi_0}.\ee Thus no Cooper pairs can tunnel through the SQUID in the adiabatic evolution for
vanishing Josephson energy, and hence the SQUID is considered to be closed. In practice, the SQUIDs cannot be perfectly closed due to asymmetry
in the tunnel junctions and finite loop inductance, which motivates us to define the minimum residual value of the Josephson energies of the
SQUIDs to be~$E_\textrm{J}^\textrm{res}$. In contrast, the maximum Josephson energy is denoted by~$E_\textrm{J}^\textrm{max}$.

\begin{figure}[tbh]
\includegraphics[width=200pt]{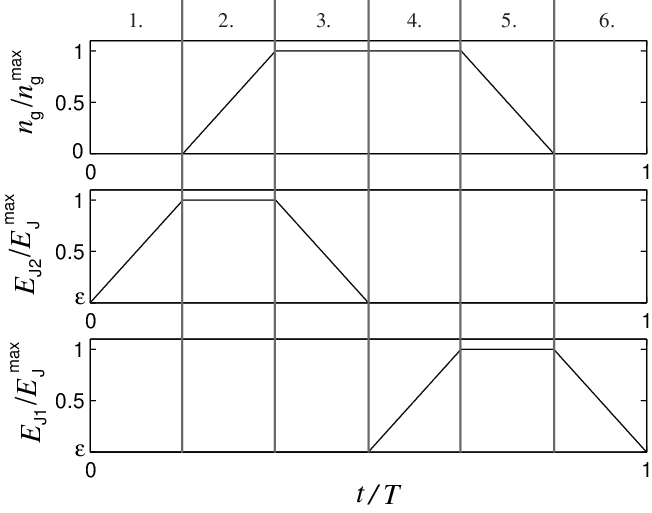}
\caption{\label{fig2} Values of the externally controlled parameters~$\eja(\Phi_1)$, $\ejb(\Phi_2)$, and~$n_\textrm{g}$ of the Hamiltonian given
by Eq.~\eqref{ham} during an example pumping cycle of the Cooper pair sluice. The Josephson energies~$\eja$ and~$\ejb$, and the gate
charge~$n_\textrm{g}$ connected to the gate voltage $V_\textrm{g}=2en_\textrm{g}/C_\textrm{g}$ are expressed in terms of their maximum
values~$E_\textrm{J}^\textrm{max}$ and~$n_\textrm{g}^\textrm{max}$, respectively. The period of one pumping cycle is denoted by~$T$. The
different phases in the pumping cycle are separated by vertical lines. For $\epsilon=E_\textrm{J}^\textrm{res}/E_\textrm{J}^\textrm{max}=0$, the
pumping cycle is referred to as ideal, and hence either one of the SQUIDs is always perfectly closed.}
\end{figure}

A typical pumping cycle of the sluice is described in Fig.~\ref{fig2}. In the following, we assume that the SQUIDs can be perfectly closed,
i.e., $E_\textrm{J}^\textrm{res}/E_\textrm{J}^\textrm{max}=0$. In the beginning of the cycle both of the SQUIDs are closed and the gate voltage
is zero implying that the ground state of the Hamiltonian given by Eq.~\eqref{ham} is an eigenstate of charge $\hat{n}$ with eigenvalue zero.
Then the second SQUID is opened by increasing~$\Phi_2$ to~$\Phi_0/2$, while the first one is kept closed. The increase of the flux~$\Phi_2$ is
executed adiabatically such that the system stays in its ground state which is, however, not an eigenstate of charge~$\hat{n}$ for
finite~$\Phi_2$. In the second step, the gate charge is brought to integer~\footnote{Although we assume~$n_\textrm{g}^\textrm{max}$ to be
integer, measurements of the pumped current~\cite{Niskanen2005} as a function of continuous~$n_\textrm{g}^\textrm{max}$ have resulted in a
step-like behavior which suggests that the pumped current is insensitive to errors in the gate charge.}~$n_\textrm{g}^\textrm{max}$, after which
the second SQUID is closed. At this middle point of the pumping cycle, the system is again in the charge eigenstate but corresponding to
eigenvalue~$n_\textrm{g}^\textrm{max}$. Since the first SQUID has been closed for the whole evolution, we conclude that
the~$n_\textrm{g}^\textrm{max}$ Cooper pairs have tunneled through the second SQUID to the island. Then we direct
the~$n_\textrm{g}^\textrm{max}$ Cooper pairs through the first SQUID by opening it, taking~$n_\textrm{g}$ to zero, and closing it.
Thus~$n_\textrm{g}^\textrm{max}$ Cooper pairs have flowed through the sluice creating an effective current of
$I_\textrm{p}=2en_\textrm{g}^\textrm{max}/T$. Although the above arguments are valid in general, we will calculate analytically the pumped
current in Sec.~\ref{sec3} for $E_\textrm{C}\gg E_\textrm{J}^\textrm{max}$ and finite~$E_\textrm{J}^\textrm{res}$. In analogy with a classical
pump, the gate voltage in the sluice corresponds to a piston and the SQUIDs correspond to valves.

The maximum Josephson energy~$E_\textrm{J}^\textrm{max}$ together with the adiabaticity criterion imposes upper bounds to the maximum pumped
current, which we consider in detail in Sec.~\ref{sec4}. In the adiabatic temporal evolution, the SQUIDs stay in the superconducting state, and
hence the average current through the sluice is given by \be \hat{I}=\frac{2e}{\hbar}\frac{\partial \hat{H}_\textrm{sl}}{\partial
\varphi}=\frac{e}{\hbar}\left(\frac{\partial \hat{H}_\textrm{sl}}{\partial \varphi_1}+\frac{\partial \hat{H}_\textrm{sl}}{\partial \varphi_2}
\right)=\frac{\hat{I}_1+\hat{I}_2}{2}.\ee The above equation yields a rather large upper bound for the effective pumped current, i.e., the maximum
critical current $I_\textrm{c}^\textrm{max}=2\pi E_\textrm{J}^\textrm{max}/\Phi_0$ of the SQUIDs.

\section{Manifestation of the Berry phase in Cooper pair pumps}\label{sec3}
Let us concentrate on the relation between the accumulated Berry phase and the pumped current. In the case of one-dimensional arrays of
Josephson junctions, the expression of the pumped charge was first obtained in Ref.~\cite{Pekola1999} and its relation to the Berry phase was
found in Ref.~\cite{Aunola2003}. Work concerning not only the Berry phase but also non-Abelian phases and their connection to the pumped charge
in rather general Josephson devices is currently in preparation~\cite{valentina}.

In Sec.~\ref{sec3a}, we consider a more general system than the sluice presented in Fig.~\ref{fig1}. Here we assume in excess of cyclic and
adiabatic temporal evolution only that the average current operator $\hat{I}=2e(\partial_\varphi \hat{H})/\hbar$ is well defined, i.e, that the
Cooper pair pump has only one input and one output. We derive the connection between the pumped charge and the Berry phase for this rather
general pump. In Sec.~\ref{sec3b}, we focus on the sluice and calculate the pumped current using a two-state approximation which is valid in the
limit $E_\textrm{C}\gg E_\textrm{J}^\textrm{max}$.

\subsection{Connection between the pumped charge and the Berry phase}\label{sec3a}
In adiabatic temporal evolution, an eigenstate of the system Hamiltonian~$\hat{H}$ is slowly varied with respect to a set of real
parameters~$\{q_k\}$, here denoted by a vector~$\mathbf{q}$. The initial state of the system at $t=t_0$ must be an
eigenstate~$\ket{m;\mathbf{q}}$ of the Hamiltonian, for which one obtains \be\label{ibasis}
\hat{H}(\mathbf{q})\ket{m;\mathbf{q}}=\varepsilon_m(\mathbf{q})\ket{m;\mathbf{q}},\ee where~$\varepsilon_m(\mathbf{q})$ is the eigenenergy of
the corresponding state. In this section, we assume that the system is very accurately in its non-degenerate eigenstate~$\ket{r;\mathbf{q}}$.
For these so-called zeroth order assumptions, the state of the system at any time instant is described by~\cite{Bransden}
\be\label{psi}\ket{\psi_0(t)}=e^{i\theta_r(t)}\ket{r;\mathbf{q}},\ee where the phase
$\theta_r(t)=\theta_{\textrm{d}r}(t)+\theta_{\textrm{g}r}(t)$ is a sum of the dynamic phase \be\label{dp}
\theta_{\textrm{d}r}=-\frac{1}{\hbar}\int_{t_0}^t\bra{r;\mathbf{q}(\tau)}\hat{H}\ket{r;\mathbf{q}(\tau)}\,\textrm{d}\tau=-\frac{1}{\hbar}\int_{t_0}^t\varepsilon_r\!\left(\mathbf{q}(t)\right)\,\textrm{d}\tau,\ee
and the geometric phase \be\label{gp}
\theta_{\textrm{g}r}=i\int_{t_0}^t\bra{r;\mathbf{q}(\tau)}\partial_\tau\ket{r;\mathbf{q}(\tau)}\,\textrm{d}\tau
=\int_{\mathbf{q}({t_0})}^{\mathbf{q}(t)}\bra{r;\mathbf{q}}\nabla_\mathbf{q}\ket{r;\mathbf{q}}\cdot\textrm{d}\mathbf{q} .\ee In cyclic temporal
evolution, the dynamic phase is proportional to the period~$T$ of the cycle, but the Berry phase, defined by \be\label{berry}
\theta_{\textrm{B}r}=i\oint_\varsigma \bra{r;\mathbf{q}}\nabla_\mathbf{q}\ket{r;\mathbf{q}}\cdot\textrm{d}\mathbf{q},\ee only depends on the
contour~$\varsigma=\{\mathbf{q}(t)|{t_0}\leq t<T+{t_0}\}$, according to which the state~$\ket{r;\mathbf{q}}$ is varied in the cycle.

To obtain the connection between the Berry phase and the pumped current~\cite{Pekola1999,valentina}, we consider the leading order corrections
to the state vector~$\ket{\psi(t)}$ in the speed of the pumping cycle~$|\dot{\mathbf{q}}|$. Hence, we express it as
\be\label{comp}\ket{\psi(t)}=\sum_{k=0}^\infty C_k(t)e^{i\theta_{k}(t)}\ket{k;\mathbf{q}} ,\ee where~$C_k(t)$ are complex numbers.
By differentiating Eq.~\eqref{comp} with respect to time and employing the Schr\"odinger equation we obtain
\be\label{ceq}\dot{C}_m\!(t)=-\sum_{k\ne m}C_k\!(t)e^{i[\theta_{k}(t)-\theta_{m}(t)]}\bra{m;\mathbf{q}}\partial_t\ket{k;\mathbf{q}}.\ee
 Since we control our system such that it remains almost exactly in the state~$\ket{r;\mathbf{q}}$, we may neglect all other terms in Eq.~\eqref{ceq} except
the one with $k=r$. Thus we obtain for $m\ne r$ the first order correction to the coefficients~$C_k$ as
\be\label{ctx}C_m(t)=-\int_{t_0}^t\bra{m;\mathbf{q}}\partial_t\ket{r;\mathbf{q}}e^{i[\theta_{r}(t')-\theta_{m}(t')]}\,\textrm{d}t'.\ee As shown
for example in Sec.~\ref{sec4}, the absolute value of the above coefficients~$C_k$ is proportional to~$|\dot{\mathbf{q}}|$ in adiabatic and
cyclic temporal evolution.

The total mean charge through the system in one pumping cycle may be written with the help of the average current operator~$\hat{I}$ as
\be\label{qtot} Q_\textrm{tot}=\int_{t_0}^{t_0+T}\bra{\psi(t)}\hat{I}\ket{\psi(t)}\,\textrm{d}t.\ee The contribution to the total charge purely
from the zeroth order term presented in Eq.~\eqref{psi} is observed to originate from the usual supercurrent and is denoted by \be\label{qs}
Q_{\textrm{s}r}=\int_{t_0}^{t_0+T}\bra{r;\mathbf{q}(t)}\hat{I}\ket{r;\mathbf{q}(t)}\,\textrm{d}t.\ee On the other hand, the pumped charge
$Q_\textrm{p}:=Q_\textrm{tot}-Q_\textrm{s}$ assumes the form \be Q_{\textrm{p}r}=2\Re\textrm{e}\int_{t_0}^{t_0+T}\left(\sum_{{k\ne r}}
C_k(t)\bra{r;\mathbf{q}(t)}\hat{I}\ket{k;\mathbf{q}(t)}e^{i[\theta_{k}(t)-\theta_{r}(t)]} \right)\textrm{d}t,\ee where we have neglected the
terms containing $|C_k|^2$ which vanish in the adiabatic limit. By inserting Eq.~\eqref{ctx} into the above equation and changing the order of
the resulting integrals we obtain \ba Q_{\textrm{p}r}=&&-2\Re\textrm{e}\sum_{k\ne r}\int_{t_0}^{t_0+T}\Big\{
\bra{k;\mathbf{q}(t')}\partial_t\ket{r;\mathbf{q}(t')}e^{i[\theta_{r}(t')-\theta_{k}(t')]}\nonumber \\ &&
\times\int_{t'}^{t_0+T}\bra{r;\mathbf{q}(t)}\hat{I}\ket{k;\mathbf{q}(t)}e^{i[\theta_{k}(t)-\theta_{r}(t)]}\,\textrm{d}t\Big\}\textrm{d}t'.\ea In
the above integral with respect to~$t$, the integrand may be expressed as \be
\frac{\textrm{d}}{\textrm{d}t}\left\{i\hbar\bra{r;\mathbf{q}(t)}\hat{I}\ket{k;\mathbf{q}(t)}e^{i[\theta_{k}(t)-\theta_{r}(t)]}/(\varepsilon_k-\varepsilon_r)\right\}+\mathcal{O}(|\dot{\mathbf{q}}|),\ee
The term~$\mathcal{O}(|\dot{\mathbf{q}}|)$ in the above equation may be dropped since it becomes negligible in the adiabatic limit compared with
the first term. Hence the pumped charge is obtained to be \ba Q_{\textrm{p}r}=&& 2\hbar\Im\textrm{m}\sum_{k\ne
r}\Big[\int_{t_0}^{t_0+T}\frac{\bra{r;\mathbf{q}(t)}\hat{I}\ket{k;\mathbf{q}(t)}\bra{k;\mathbf{q}(t)}\partial_t\ket{r;\mathbf{q}(t)}}{\varepsilon_r-\varepsilon_k}\,\textrm{d}t
\nonumber \\
&&+\frac{C_k(t_0+T)\bra{r;\mathbf{q}(t_0+T)}\hat{I}\ket{k;\mathbf{q}(t_0+T)}e^{i[\theta_k(t_0+T)-\theta_r(t_0+T)]}}{\varepsilon_r-\varepsilon_k}\Big].
\ea Since the excitation amplitudes $C_k$ tend to zero in the adiabatic limit, we obtain the known equation~\cite{Pekola1999} for the pumped
charge \be\label{qp} Q_{\textrm{p}r}=2\hbar\Im\textrm{m}\oint_\varsigma\sum_{k\ne
r}\frac{\bra{r;\mathbf{q}}\hat{I}\ket{k;\mathbf{q}}}{\varepsilon_r-\varepsilon_k}\bra{k;\mathbf{q}}\nabla_\mathbf{q}\ket{r;\mathbf{q}}\cdot\textrm{d}\mathbf{q},\ee
which only depends on the contour~$\varsigma$, according to which the state~$\ket{r;\mathbf{q}}$ traverses, not the speed~$|\dot{\mathbf{q}}|$
of the pumping cycle. Although the above equation contains an integral over a closed contour, it is also valid for open contours. Furthermore,
the matrix elements of the average current operator~$\hat{I}=2e[\partial_\varphi,\hat{H}]/\hbar$ can be expressed as
\be{\bra{r;\mathbf{q}}\hat{I}\ket{k;\mathbf{q}}}=\frac{2e(\varepsilon_k-\varepsilon_r)}{\hbar}{\bra{r;\mathbf{q}}\partial_\varphi\ket{k;\mathbf{q}}}
,\ee which yields together with Eq.~\eqref{qp} \be\label{qpxx}
Q_{\textrm{p}r}=-4e\Im\textrm{m}\oint_\varsigma\bra{r;\mathbf{q}}\partial_\varphi\nabla_\mathbf{q}\ket{r;\mathbf{q}}\cdot\textrm{d}\mathbf{q}.\ee
In the above derivation, we have employed the fact that
$\bra{r;\mathbf{q}}\partial_\varphi\ket{r;\mathbf{q}}\bra{r;\mathbf{q}}\nabla_\mathbf{q}\ket{r;\mathbf{q}}$ is real and that the basis
$\{\ket{k;\mathbf{q}}\}$ is normalized spaning the whole configuration space.

Equation~\eqref{qpxx} for the pumped charge resembles Eq.~\eqref{berry} except for the operator~$\partial_{\varphi}$. This observation motivates
us to study the derivative of the Berry phase with respect to the phase difference across the sluice~$\varphi$ which is fixed by the total
magnetic flux~$\Phi$ as in Eq.~\eqref{totphase}, and hence may be treated as a scalar variable. Hence, Eq.~\eqref{berry} yields
\be\label{turha2}
\partial_\varphi\theta_{\textrm{B}r}=i\oint_\varsigma\left[\left(\partial_\varphi\ket{r;\mathbf{q}}\right)^\dagger\nabla_\mathbf{q}\ket{r;\mathbf{q}}+\bra{r;\mathbf{q}}\partial_\varphi\nabla_\mathbf{q}\ket{r;\mathbf{q}}\right]\cdot\textrm{d}\mathbf{q}.\ee
The first term in the right side of Eq.~\eqref{turha2} may be written in a form \be
\left(\partial_\varphi\ket{r;\mathbf{q}}\right)^\dagger\nabla_\mathbf{q}\ket{r;\mathbf{q}}=\nabla_\mathbf{q}\left[\left(\partial_\varphi\ket{r;\mathbf{q}}\right)^\dagger\ket{r;\mathbf{q}}\right]-\left(\nabla_\mathbf{q}\partial_\varphi\ket{r;\mathbf{q}}\right)^\dagger\ket{r;\mathbf{q}}.\ee
The integral of the term~$\nabla_\mathbf{q}\left[\left(\partial_\varphi\ket{r;\mathbf{q}}\right)^\dagger\ket{r;\mathbf{q}}\right]$ over a closed
contour~$\varsigma$ vanishes by the Stokes theorem. Thus Eq.~\eqref{turha2} may be recast into
\be\label{phib}\partial_\varphi\theta_{\textrm{B}r}=-2\Im\textrm{m}\oint_\varsigma\bra{r;\mathbf{q}}\partial_\varphi\nabla_\mathbf{q}\ket{r;\mathbf{q}}\cdot\textrm{d}\mathbf{q},\ee
and hence the comparison of Eqs.~\eqref{qpxx} and~\eqref{phib} yields the relation between the Berry phase and the pumped charge
\be\label{connection} Q_\textrm{p}=2e\partial_\varphi\theta_\textrm{B}.\ee The interpretation of $Q_\textrm{p}$ is the average pumped charge
through the device, since it arises from the average current operator $\hat{I}$. Furthermore, if the pump consists of a linear chain of islands
separated by SQUIDs, the charge conservation and cyclic temporal evolution assures that the pumped charge through all the SQUIDs separately
equal to~$Q_\textrm{p}$~\cite{Pekola1999}. Since the Berry phase is not directly observable in a one-dimensional subspace of any quantum system,
only the derivative of the Berry phase with respect to the total phase difference appears in Eq.~\eqref{connection}. We do not regard this
property to be a limitation in measuring the Berry phase in Josephson circuits, but rather equate it with the fact that generally the Berry
phase is detected as a phase difference between two orthogonal quantum states. In contrast to the Berry phase, the charge corresponding to the
dynamic phase is related to the usual supercurrent through the device as \be\label{dyn} Q_\textrm{s}=-2e\partial_\varphi\theta_\textrm{d},\ee
which is obtained by differentiating Eq.~\eqref{dp} with respect to~$\varphi$ and comparing the result with Eq.~\eqref{qs}.

\subsection{Berry phase in the Cooper pair sluice}\label{sec3b}
Assuming $E_\textrm{C}\gg E_\textrm{J}^\textrm{max}$ and $n_\textrm{g}\in[0,1]$, the state of the Cooper pair sluice can be described by two
eigenstates~$\ket{0}$ and~$\ket{1}$ of the charge operator $\hat{n}$ corresponding to eigenvalues~$0$ and~$1$, respectively. We employ this
approximation to demonstrate the results of Sec.~\ref{sec3a} for the sluice up to linear terms in
\be\label{delta}\delta:=\frac{E_\textrm{J}^\textrm{max}}{E_\textrm{C}}.\ee In the discussion of Sec.~\ref{sec2}, we found that the pumped charge
is~$2en_\textrm{g}^\textrm{max}$ in an ideal pumping cycle shown in Fig.~\ref{fig2} for $E_\textrm{J}^\textrm{res}/E_\textrm{J}^\textrm{max}=0$.
In contrast, we consider here SQUIDs with finite \be\label{epsilon}\epsilon:=\frac{E_\textrm{J}^\textrm{res}}{E_\textrm{J}^\textrm{max}}.\ee We
will, however, assume that the residual Josephson energy is small, and hence study its effect only up to linear terms in~$\epsilon$.

From the expression~$\hat{n}=-i\partial_\phi$ of the number operator, we observe that \be\label{ephi} e^{\pm i\phi}\ket{n}=\ket{n\pm 1},\ee
where~$\ket{n}$ is an eigenstate of the number operator~$\hat{n}$ with an eigenvalue~$n$. Thus we decompose the trigonometric functions of
Eq.~\eqref{ham} into exponential functions and obtain the approximate two-state Hamiltonian in the basis $\{\ket{0},\ket{1}\}$ as
\be\label{aham}\hat{H}_\textrm{2s}=\begin{pmatrix}E_\textrm{C}n_\textrm{g}^2 &
-\frac{\eja}{2}e^{-i\varphi/2}-\frac{\ejb}{2}e^{i\varphi/2} \\
-\frac{\eja}{2}e^{i\varphi/2}-\frac{\ejb}{2}e^{-i\varphi/2} & E_\textrm{C}(1-n_\textrm{g})^2\end{pmatrix}.\ee
The ground state~$\ket{\textrm{g}}$ and the excited state~$\ket{\textrm{e}}$ are obtained by diagonalizing this Hamiltonian, and can be
expressed as \ba\label{states} \ket{\textrm{g}} = e^{i\gamma}a|0\rangle+b|1\rangle, \nonumber \\ \ket{\textrm{e}} =
e^{i\gamma}b|0\rangle-a|1\rangle, \ea where the amplitudes~$a$ and~$b$ are positive real numbers satisfying $a^2+b^2=1$ and~$\gamma$ is the
relative phase difference of the charge states. The overall constant phase of the eigenstates is physically unobservable and can thus be chosen
as in Eq.~\eqref{states}. The amplitudes~$a$ and~$b$ are obtained to be \be \label{ab} a^2 = 1 - b^2 = \frac {1}{2}[1- \frac {\eta}{\sqrt
{\eta^2 + (E_{12}/E_{\rm C})^2}}],\ee where we have defined~$\eta=n_\textrm{g}-1/2$ and~$E_{12}= \frac {1}{2}\sqrt {\eja^2 + \ejb^2 + 2
\eja\ejb\cos \varphi}$. Moreover, the phase difference~$\gamma$ assumes the form \be\label{gamma} \gamma = \arctan \left(\frac{\ejb - \eja}{\eja
+ \ejb} \tan\frac{\varphi}{2}\right). \ee

Our aim is to compute the pumped charge utilizing Eq.~\eqref{qp}, and compare it with the derivative of the Berry phase with respect to the
phase difference~$\varphi$. In this two-state approximation, the sum in Eq.~\eqref{qp} consists of only one term \be\label{qp2s}Q_\textrm{p2s}=
2\hbar\Im\textrm{m}\oint_\varsigma\frac{\bra{\textrm{g};\mathbf{q}}\hat{I}_2\ket{\textrm{e};\mathbf{q}}}{\varepsilon_\textrm{g}-\varepsilon_\textrm{e}}\bra{\textrm{e};\mathbf{q}}\nabla_\mathbf{q}\ket{\textrm{g};\mathbf{q}}\cdot\textrm{d}\mathbf{q},\ee
where we have employed the fact that all the current operators~$\hat{I}$, $\hat{I}_1$, and~$\hat{I}_2$ result in equal pumped charges due to the
charge conservation. The symbols~$\varepsilon_\textrm{g}$ and~$\varepsilon_\textrm{e}$ denote the eigenvalues of the
Hamiltonian~$\hat{H}_\textrm{2s}$ corresponding to the states~$\ket{\textrm{g};\mathbf{q}}$ and~$\ket{\textrm{e};\mathbf{q}}$, respectively. The
parameters~$\mathbf{q}$ are chosen to be $\mathbf{q}=(a^2,\gamma)$. With the help of Eqs.~\eqref{iks} and~\eqref{ephi}, the matrix element of
the current operator~$\hat{I}_2$ in Eq.~\eqref{qp2s} is expressed as
\be\label{turha3}\bra{\textrm{g};\mathbf{q}}\hat{I}_2\ket{\textrm{e};\mathbf{q}}=\frac{ie}{\hbar}
\ejb[e^{i(\gamma-\varphi/2)}b^2+e^{-i(\gamma-\varphi/2)}a^2].\ee We write the gradient part in Eq.~\eqref{qp2s} as
\be\label{turha4}\bra{\textrm{e};\mathbf{q}}\nabla_\mathbf{q}\ket{\textrm{g};\mathbf{q}}\cdot\textrm{d}\mathbf{q}=iab\,\textrm{d}\gamma+\frac{\textrm{d}(a^2)}{2ab},\ee
and the energy difference as \be\label{turha5}\varepsilon_\textrm{e}-\varepsilon_\textrm{g}=2E_\textrm{C} \sqrt{E_\textrm{C}^2\,\eta^2 +
E_{12}^2}.\ee By a substitution of Eqs.~\eqref{turha3},~\eqref{turha4}, and~\eqref{turha5} into Eq.~\eqref{qp2s} we obtain
\be\label{qk2sx}Q_\textrm{p2s}= -2e \Re\mbox{e}\oint_\varsigma \frac{\ejb}{E_{12}} [e^{i(\gamma-\varphi/2)}b^2+e^{-i(\gamma-\varphi/2)}a^2]
[i\frac{E_{12}^2\,\textrm{d}\gamma}{4 (E_{\rm C}^2 \eta^2 + E_{12}^2)}+\frac{1}{2}\,\textrm{d}(a^2)].\ee Since the part
containing~$\textrm{d}\gamma$ in the above integral vanishes as $\mathcal{O}(\delta^2)$, it is neglected in this approximation. With the same
argument we neglect the terms containing~$\textrm{d}(a^2)$ for the parts of the contour~$\varsigma$ in which~$\eja$ or~$\ejb$ is varied.
Furthermore, we employ the identity $b^2=1-a^2$ to Eq.~\eqref{qk2sx} and obtain \be\label{qp2sx} Q_\textrm{p2s}=-2e\int
\frac{\ejb}{2E_{12}}\cos(\gamma-\varphi/2)\,\textrm{d}(a^2) 
+\mathcal{O}(\delta^2),\ee where the integration is over the second and fifth pumping phase, see Fig.~\ref{fig2}.
For the second pumping phase, $\ejb=E_\textrm{J}^\textrm{max}$, and up to linear order in $\epsilon$ we have $E_{12}=
{E_\textrm{J}^\textrm{max}}(1+ \epsilon\cos\varphi)/2$ and $\cos(\gamma-\varphi/2)=1$. Furthermore, $a^2$ varies from $1-\mathcal{O}(\delta^2)$
to $\mathcal{O}(\delta^2)$, and hence the contribution to the pumped charge from this domain is $2e(1-\epsilon\cos\varphi)$. For the fifth
pumping phase, $\ejb=E_\textrm{J}^\textrm{res}$, $E_{12}\approx {E_\textrm{J}^\textrm{max}}(1+ \epsilon\cos\varphi)/2$, and
$\cos(\gamma-\varphi/2)\approx\cos(\varphi)$. In this part of the cycle $a^2$ varies from $\mathcal{O}(\delta^2)$ to $1-\mathcal{O}(\delta^2)$,
and hence the contribution to the pumped charge up to linear order in $\epsilon$ is given by $-2e \epsilon\cos\varphi$. Thus the pumped charge
of the sluice in the adiabatic cycle of Fig. \ref{fig2} assumes the form \be\label{qp2sf}Q_\textrm{p2s}=2e\left(
1-2\epsilon\cos\varphi\right)+\mathcal{O}(\delta^2)+\mathcal{O}(\epsilon^2).\ee For vanishing residual Josephson energy, the above equation
reduces to the general result for ideal pumping cycle $Q_\textrm{p}=2en_\textrm{g}^\textrm{max}$, where the number of Cooper pairs pumped in one
cycle $n_\textrm{g}^\textrm{max}$ is unity.

To show the relation between the pumped charge and the Berry phase, we employ Eq.~\eqref{berry} and write
\be\label{turha5x}\theta_\textrm{B2s}=i\oint_\varsigma
\bra{\textrm{g};\mathbf{q}}\nabla_\mathbf{q}\ket{\textrm{g};\mathbf{q}}\cdot\textrm{d}\mathbf{q}=-\oint_\varsigma a^2\,\textrm{d}\gamma ,\ee
where the vanishing integral involving the differential~$\textrm{d}(a^2)$ is not shown. Since the parameter~$\gamma$ does not depend on the gate
charge, the Berry phase is accumulated only when the Josephson energies are changed in the cycle of Fig.~\ref{fig2}. Since
$a^2=\mathcal{O}(\delta^2)$ for $\eta=1/2$, the only non-negligible contribution to the integral in Eq.~\eqref{turha5x} is obtained in the first
and last pumping phase, for which $a^2=1+\mathcal{O}(\delta^2)$. Thus the Berry phase can be expressed as
\be\label{turha6}\theta_\textrm{B2s}=\varphi-2\epsilon\sin(\varphi)+\mathcal{O}(\delta^2)+\mathcal{O}(\epsilon^2). \ee Indeed,
Eqs.~\eqref{qp2sf} and~\eqref{turha6} demonstrate that \be {Q}_\textrm{P2s} = 2e\partial_\varphi \theta_\textrm{B2s}, \ee which demonstrates for
the sluice the validity of the general result given by Eq.~\eqref{connection}.


\section{Adiabaticity criterion for the Cooper pair sluice}\label{sec4}
In the above sections, we have assumed that the sluice is more or less perfectly in its instantaneous ground state during the temporal
evolution. However, this assumption is strictly speaking valid only if the changes in the parameters~$\mathbf{q}$ of the system Hamiltonian are
infinitely slow. On the other hand, we wish to repeat the pumping cycle as fast as possible for a given~$n_\textrm{g}^\textrm{max}$ to create a
detectable current. Thus a quantitative study of the pumping errors due to finite pumping rate is required. In this section, we derive upper
bounds for the population of the excited states of the sluice in the spirit of adiabatic theorem. Moreover, we apply the results to the case
$\delta\ll 1$ for which the two-state approximation is valid, and to the case $\delta\gg 1$ for which we employ a perturbed harmonic oscillator
approach. In these two cases, we also derive the maximum pumped current for small pumping error rate.

As in Sec.~\ref{sec3a}, we study a rather general quantum system described by the Hamiltonian~$\hat{H}(\mathbf{q})$, parameters~$\mathbf{q}$,
and the instantaneous eigenbasis~$\{\ket{m;\mathbf{q}}\}$ defined in Eq.~\eqref{ibasis}. We consider a representation of the state of the system
given by Eq.~\eqref{comp}. Since the aim of this section is to show how we can retain the system in its ground state, i.e., $|C_0(t)|^2\approx
1$, we may employ Eq.~\eqref{ctx} with $r=0$ to describe the time evolution of the excitation coefficients~$C_k$. It follows from
Eq.~\eqref{ibasis} that
\be\label{turha7}\bra{m;\mathbf{q}}\partial_t\ket{k;\mathbf{q}}=\frac{\bra{m;\mathbf{q}}(\partial_t\hat{H})\ket{k;\mathbf{q}}}{\varepsilon_k(t)-\varepsilon_m(t)}
.\ee By insertion of Eq.~\eqref{turha7} into Eq.~\eqref{ctx} we obtain
\be\label{ct}C_m(t)=\int_{t_0}^t\frac{\bra{m;\mathbf{q}}(\partial_{t'}\hat{H})\ket{0;\mathbf{q}}}{\varepsilon_m(t')-\varepsilon_0(t')}e^{i[\theta_{m}(t')-\theta_{0}(t')]}\,\textrm{d}t'.\ee
Let us assume that we adjust the speed of the temporal evolution such that
$\bra{m;\mathbf{q}}(\partial_{t}\hat{H})\ket{0;\mathbf{q}}/[\varepsilon_m(t)-\varepsilon_0(t)]^2$ is constant in time. Furthermore, if we
neglect the geometric part of the phase and assume that the energy separation is time independent, we obtain an estimate
\be\label{exprob}|C_m(t)|\leq
2\hbar\frac{|\bra{m;\mathbf{q}}(\partial_{t}\hat{H})\ket{0;\mathbf{q}}|}{[\varepsilon_m(t)-\varepsilon_0(t)]^2}.\ee Although the above
assumptions may be considered to be crude in general, they turn out to be accurate for a sluice Hamiltonian with at least one of the Josephson
energies much larger than the Coulomb energy~$E_\textrm{C}$.

Without any other assumption about the temporal evolution of the system except that it is cyclic and nearly adiabatic, Eq.~\eqref{ct} yields
\be\label{ctc}C_m(NT+t_0)=C_m(T+t_0)\frac{1-\nu_m^{N+1}}{1-\nu_m},\quad\textrm{for }\nu_m\neq 1 ,\ee where~$N$ is a positive integer and
$\nu_m=e^{i[\theta_{m}(T)-\theta_{0}(T)]}$. Since the dynamic phase depends explicitly on the period~$T$ and the geometric phase only on the
contour~$\varsigma$, according to which the parameters~$\mathbf{q}$ vary under the cyclic temporal evolution, the period~$T$ can be chosen for
example such that $\nu_m=i$ for a fixed~$m$, which implies $|C_m(NT)|<\sqrt{2}|C_m(T)|$. Thus the probability for the sluice to be in the
excited state~$\ket{\mathbf{q};m}$ is bounded. The above discussion is, however, strictly valid only for the $m$th state, whereas $\nu_k$ can
generally be very close to unity for $k\ne m$. In the extreme case $\nu_k=1$, the amplitude of the $k$th state increases linearly in time, and
hence the adiabaticity is lost if there is no dissipation in the system. To take the dissipation into account, we estimate from Eq.~\eqref{ct}
that
\be\label{cdt}\dot{C}_m(t)\leq\frac{|\bra{m;\mathbf{q}}(\partial_{t}\hat{H})\ket{0;\mathbf{q}}|}{\varepsilon_m(t)-\varepsilon_0(t)}\leq\max_{0<t<T}\frac{|\bra{m;\mathbf{q}}(\partial_{t}\hat{H})\ket{0;\mathbf{q}}|}{\varepsilon_m(t)-\varepsilon_0(t)}:=\kappa_m.\ee
For the excitation probability $P_m(t)=|C_m(t)|^2$, Eq.~\eqref{cdt} and addition of relaxation to the system yields
\be\label{pdt}\dot{P}_m\leq4\kappa_m\sqrt{P_m(t)}-\Gamma_mP_m(t),\ee where~$\Gamma_m$ is the relaxation rate of the state~$\ket{\mathbf{q};m}$.
Equation~\eqref{pdt} yields a bound $(4\kappa_m/\Gamma_m)^2$ for the excitation probability~$P_m(t)$, and hence
\be\label{cb}|C_m(t)|\leq\frac{4\kappa_m}{\Gamma_m}.\ee From Eq.~\eqref{cb} we can conclude that if the relaxation is dominant, i.e.,
$\Gamma_m\geq 2(\varepsilon_m-\varepsilon_0)/\hbar$, bound in Eq.~\eqref{exprob} is valid even though the assumption that the energy difference
$\varepsilon_m-\varepsilon_0$ is independent of time fails. Moreover, since~$\kappa_m$ tends to decrease rapidly with increasing~$m$, it may be
argued that the excitation amplitudes of the high-lying states are suppressed by relaxation, and hence we have to take only the low-lying states
into account if we consider an isolated quantum system as a model for the physical system.

The errors to the pumped current arise from the leakage current due to finite residual Josephson energy~$E_\textrm{J}^\textrm{res}$ and from the
non-adiabaticity of the temporal evolution which we consider here. Since the pumped current is determined by~$2en_\textrm{g}^\textrm{max}/T$,
the optimal way to pump is to choose the integer~$n_\textrm{g}^\textrm{max}$ to be as large as possible to decrease the contribution to the
period~$T$ from opening and closing the SQUIDs, see Fig.~\ref{fig2}. Thus the maximum pumped current is determined by the maximum value
for~$\dot{n}_\textrm{g}$ limited by the criteria of Eqs.~\eqref{exprob} and~\eqref{cb} when one of the SQUIDs is fully open and the other one is
closed. Since the energy difference $\varepsilon_\textrm{e}-\varepsilon_\textrm{g}$ depends strongly on $n_\textrm{g}$ in the two state
approximation valid for $\delta\ll 1$, the estimate in Eq.~\eqref{exprob} does not hold. In contrast, we adjust~$\dot{n}_\textrm{g}$ for all~$t$
such that
\be\label{turha8}\frac{|\bra{\textrm{e};\mathbf{q}}(\partial_{t}\hat{H}_\textrm{sl})\ket{\textrm{g};\mathbf{q}}|}{\varepsilon_\textrm{e}(t)-\varepsilon_\textrm{g}(t)}=\kappa_\textrm{e},\ee
where $\partial_t \hat{H}_\textrm{sl}=2E_\textrm{C}\dot{n}_\textrm{g}(n_\textrm{g}-\hat{n})$. Substitution of Eqs.~\eqref{states}
and~\eqref{turha5} into Eq.~\eqref{turha8} results in a differential equation for the gate charge \be\label{dng2s}
\dot{n}_\textrm{g}=\frac{2\kappa_\textrm{e}}{\delta}[(n_\textrm{g}-1/2)^2+\delta^2],\ee which implies together with the initial condition
$n_\textrm{g}(0)=0$ that \be\label{ngt2s} n_\textrm{g}(t)=\frac{1}{2}+{\delta}\tan\left[2\kappa_\textrm{e}t-\arctan(\delta^{-1}/2)\right]. \ee
The above equation yields the period~$T=(\pi/2-2\delta)/\kappa_\textrm{e}+\mathcal{O}(\delta^3)$ and the pumped current \be\label{i2s}
I_\textrm{p2s}=-{2e\kappa_\textrm{e}}\left(\frac{2}{\pi}+\frac{8\delta}{\pi^2}\right)+\mathcal{O}(\delta^2),\ee where~$\kappa_\textrm{e}$ is to
be chosen such that $\kappa_\textrm{e}\ll \Gamma_\textrm{e}/4$, see Eq.~\eqref{cb}. Although a high damping rate is desirable in light of the
adiabaticity, it requires a strong coupling of the system to its environment. The strong coupling has disadvantages, since it may introduce
noise to the control parameters of the Hamiltonian, and hence result in pumping errors.

\begin{figure}[tbh]
\includegraphics[width=200pt]{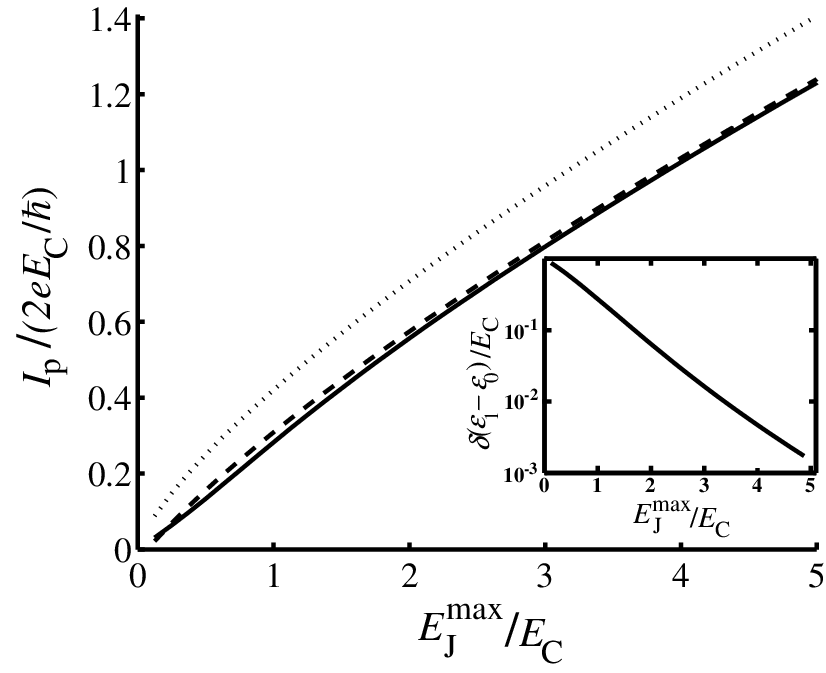}
\caption{\label{fig3} Pumped current as a function of the maximum Josephson energy of the SQUIDs~$E_\textrm{J}^\textrm{max}$ obtained from
Eq.~\eqref{ihoap} (dotted line), Eq.~\eqref{ipho} (dashed), and full numerical solution of the system (solid line). The inset shows the maximum
relative variation of the energy difference $\varepsilon_1-\varepsilon_0$ with respect to~$n_\textrm{g}$ as a function of the maximum Josephson
energy of the SQUIDs.}
\end{figure}

Equation~\eqref{i2s} shows that the pumped current increases with $\delta=E_\textrm{J}^\textrm{max}/E_\textrm{C}$. This observation motivates us
to work in the limit $\delta\gg 1$ for which the phase states are accurate eigenstates of the Hamiltonian in Eq.~\eqref{ham}. Since the ground
state of the system is localized at the minimum of the potential, we may approximate the Hamiltonian~$\hat{H}_\textrm{sl}$ with a Hamiltonian of
a perturbed harmonic oscillator as
\be\label{hamapp}\hat{H}_\textrm{pho}=-E_\textrm{C}\partial_{{\phi\,'}}^2+\frac{\ejb}{2}(\phi\,')^2-\ejb+\hat{H}_\textrm{p},\ee where
$\phi\,'=\phi-\varphi/2$, the perturbation Hamiltonian assumes the form $\hat{H}_\textrm{p}=-\ejb[\cos(\phi\,')-1+(\phi\,')^2/2] $, and we have
assumed for simplicity that $\eja=0$. Furthermore, we have neglected the contribution from~$n_\textrm{g}$ since it can be taken into account in
the unperturbed case by multiplying the wave function by $e^{in_\textrm{g}\phi\,'}$ which does not introduce energy shifts or changes into the
matrix elements~$\bra{k;\mathbf{q}}(\partial_{n_\textrm{g}}\hat{H}_\textrm{sl})\ket{m;\mathbf{q}}$. By solving the eigenstates of the
unperturbed Hamiltonian $\hat{H}_\textrm{pho}-\hat{H}_\textrm{p}$ and applying perturbation theory up to any order beyond the first it can be
shown that \ba\label{eneh} \frac{\varepsilon_m-\varepsilon_0}{E_\textrm{C}}=\sqrt{2\delta }m-2^{-4}(2m^2+2m)+\mathcal{O}(\delta^{-1/2}),\ea and
\be\label{matelho}\frac{\bra{m;\mathbf{q}}(\partial_{n_\textrm{g}}\hat{H}_\textrm{sl})\ket{0;\mathbf{q}}}{E_\textrm{C}}=\left\{\begin{array}{ll}
i2^{1/4}(\delta^{1/4}-2^{-7/2}\delta^{-1/4})+\mathcal{O}(\delta^{-3/4}) & \textrm{, for }m=1 \\ \mathcal{O}(\delta^{-1/4}) & \textrm{, for }m\ne
1 \end{array} \right. .\ee In fact the eigenstates of the unperturbed Hamiltonian are also the eigenstates of the perturbed Hamiltonian up
to~$\mathcal{O}(\delta^{-1/2})$, which justifies our assumption that~$n_\textrm{g}$ may be neglected also in the perturbed case. We showed in
the above discussion that the assumptions leading to the bound in Eq.~\eqref{exprob} are valid here. Substitution of Eqs.~\eqref{eneh}
and~\eqref{matelho} into Eq.~\eqref{exprob} for $m=1$ yields
\be\label{exprobho}|C_1(t)|\leq\frac{\hbar\dot{n}_\textrm{g}}{E_\textrm{C}}\left[2^{1/4}\delta^{-3/4}+3\times
2^{-13/4}\delta^{-5/4}+\mathcal{O}(\delta^{-7/4})\right]:=\lambda.\ee Hence, the pumped current in this limit $E_\textrm{J}^\textrm{max}\gg
E_\textrm{C}$ is obtained from \be\label{ipho}
I_\textrm{pho}=e\dot{n}_\textrm{g}=\frac{eE_\textrm{C}\lambda}{\hbar}\left[2^{-1/4}\delta^{3/4}-3\times
2^{-15/4}\delta^{1/4}+\mathcal{O}(\delta^{-1/4})\right],\ee where we choose $\lambda\ll 1$ to assure that the system is accurately in its ground
state. The leading contribution to the current may be expressed as \be\label{ihoap} I_\textrm{pho}\approx \frac{2^{-1/4}e\lambda
E_\textrm{J2}^{3/4}E_\textrm{C}^{1/4}}{\hbar},\ee where the Josephson energy~$\ejb=E_\textrm{J}^\textrm{max}$. To justify the accuracy of above
approximations, we show in Fig.~\ref{fig3} the estimates for the pumped current obtained from Eqs.~\eqref{ihoap} and~\eqref{ipho} compared with
a numerical solution, for which the spectrum of the Hamiltonian~$\hat{H}_\textrm{sl}$ was calculated, and the gate charge~$n_\textrm{g}$ was
ramped up employing Eq.~\eqref{exprob} and the condition $|C_m(t)|<\lambda$ for all~$m$. On the other hand, the inset in Fig.~\ref{fig3} shows
that dependence of the energy difference $\varepsilon_1-\varepsilon_0$ on~$n_\textrm{g}$, and hence time, weakens with increasing~$\delta$,
which renders Eq.~\eqref{exprob} to be valid. Thus we conclude that Eq.~\eqref{ipho} yields an accurate analytic expression for the pumped
current in a Cooper pair sluice for $E_\textrm{J}^\textrm{max}\gtrsim E_\textrm{C}$.

The calculation of the detailed dependence of the pumping errors~$\delta I_\textrm{p}$ on the parameter~$\lambda$ is beyond the scope of this
paper. Instead, we will employ an approximation \be \frac{\delta I_\textrm{p}}{I_\textrm{p}}\approx \lambda^2.\ee Thus we only give order of
magnitude estimates for the maximum pumped current with a given error rate.

Figure~\ref{fig3} shows that to maximize the pumped current, we should ultimately maximize~$E_\textrm{J}^\textrm{max}$. Apart from choosing the
material of the superconductor such that its energy gap is as large as possible and using very transparent tunnel junctions, the only way to
increase the maximum Josephson energy of the SQUIDs is to increase the surface area of the tunnel junctions. However, the junction capacitance
increases with the surface area, and hence the Coulomb energy~$E_\textrm{C}$ decreases. On the other hand, the Coulomb energy should be large
compared with the temperature for the finite temperature effects to be minimal. Thus the attainable base temperature
limits~$E_\textrm{J}^\textrm{max}$. For standard experimental parameters, i.e., relatively transparent aluminum junctions,
$E_\textrm{C}/k_\textrm{B}=1$~K, and for $\lambda=0.01$ we obtain currents of the order of a nanoampere. The pumping errors are not critical for
observation of the Berry phase, and hence the parameter~$\lambda$ can be chosen to be, e.g., $0.1$ which implies lower than one percent
population in the first excited state and a pumped current of the order of $10$~nA. On the other hand, if the sluice is used to define the
current standard, relative pumping errors should be less than $10^{-7}$ implying  $\lambda<3.1\times 10^{-4}$. In this case, the standard
parameters still yield~$I_\textrm{pho}$ to be of the order of~$10$~pA. For superconducting niobium, the gap in the quasiparticle energy spectrum
is an order of magnitude higher than for aluminum leading to about six times larger pumped currents than for aluminum devices.

\section{Measurement scheme for the pumped current}\label{sec5}
\begin{figure}[tbh]
\includegraphics[width=200pt]{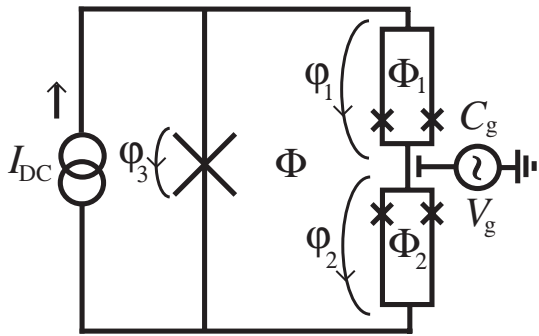}
\caption{\label{fig4} Circuit diagram for the measurement of the pumped current through the Cooper pair sluice. The third tunnel junction and a DC
current source with current~$I_\textrm{DC}$ is installed in parallel with the sluice shown in Fig.~\ref{fig1}. The phase difference of the third
junction is denoted by~$\varphi_3$.} 
\end{figure}

To observe the pumped current of a phase biased Cooper pair sluice, we install a third tunnel junction and a DC current source with
current~$I_\textrm{DC}$ in parallel with the sluice, as shown in Fig.~\ref{fig4}. The applied current~$I_\textrm{DC}$ is assumed to be smaller than the
critical current of the system, and hence there is no voltage over the third junction. In addition, the capacitance of the third junction is
assumed to be large enough such that its charging energy is negligible. Thus the third junction is essentially in such an eigenstate of its
phase operator~$\varphi_3$ that the free energy of the system is minimized. We write the free energy of the circuit shown in Fig.~\ref{fig4} as
\be \mathcal{G}=\hat{H}_\textrm{sl}-\ejc\cos\varphi_3-\frac{\hbar I_\textrm{DC}}{2e}\varphi_3, \ee where the second and the third term
correspond to the Josephson energy of the third junction and to the work done by the current source, respectively. Since the system is in its
ground state in the adiabatic temporal evolution, the free energy must achieve its minimum at~$\varphi_3$, leading to
\be\label{ming}\partial_{\varphi_3}\mathcal{G}=\ejc\sin{\varphi_3}+\frac{\hbar}{2e}(I_\textrm{sl}-I_\textrm{DC})=0, \ee where 
$I_\textrm{sl}=(\eja\sin\varphi_1+\ejb\sin\varphi_2)/2$ denotes the total average supercurrent through the sluice, and we have used the relation
$2\pi\Phi/\Phi_0=\varphi_1+\varphi_2-\varphi_3$. In Eq.~\eqref{ming}, we did not, however, include the effective current
$I_\textrm{p}=2en_\textrm{g}^\textrm{max}/T$ through the third junction arising from the pumping. This current has to be subtracted from the
applied DC current and thus we obtain a relation \be\label{idc}
\tilde{I}:=I_\textrm{DC}-I_\textrm{p}=\frac{2e\ejc}{\hbar}\sin\varphi_3+I_\textrm{sl}.\ee Equation~\eqref{idc} shows that the pumping of Cooper
pairs alters the current through the system, which can be measured as a shift in the critical current of the device.

In the case of ideal SQUIDs, for which $\epsilon=E_\textrm{J}^\textrm{res}/E_\textrm{J}^\textrm{max}=0$, it follows from Hamiltonian of
Eq.~\eqref{hamapp} that the pumped current is independent of the total phase difference~$\varphi$. Thus we obtain from Eq.~\eqref{idc}
\be\label{idcx}\tilde{I}=\frac{2e\ejc}{\hbar}\sin\varphi_3,\ee which is equivalent to the case where the sluice is replaced by an ideal current
source. Since the pumping shifts the effective current~$\tilde{I}$ through the third junction by $I_\textrm{p}$, the pumping can be observed by
switching measurements, i.e., by applying DC current pulses of duration~$\tau$ and different magnitudes~$I_\textrm{DC}$ while monitoring the
voltage across the device. If the device switches to the normal conducting state, a non-zero voltage is observed.

\begin{figure}[tbh]
\includegraphics[width=200pt]{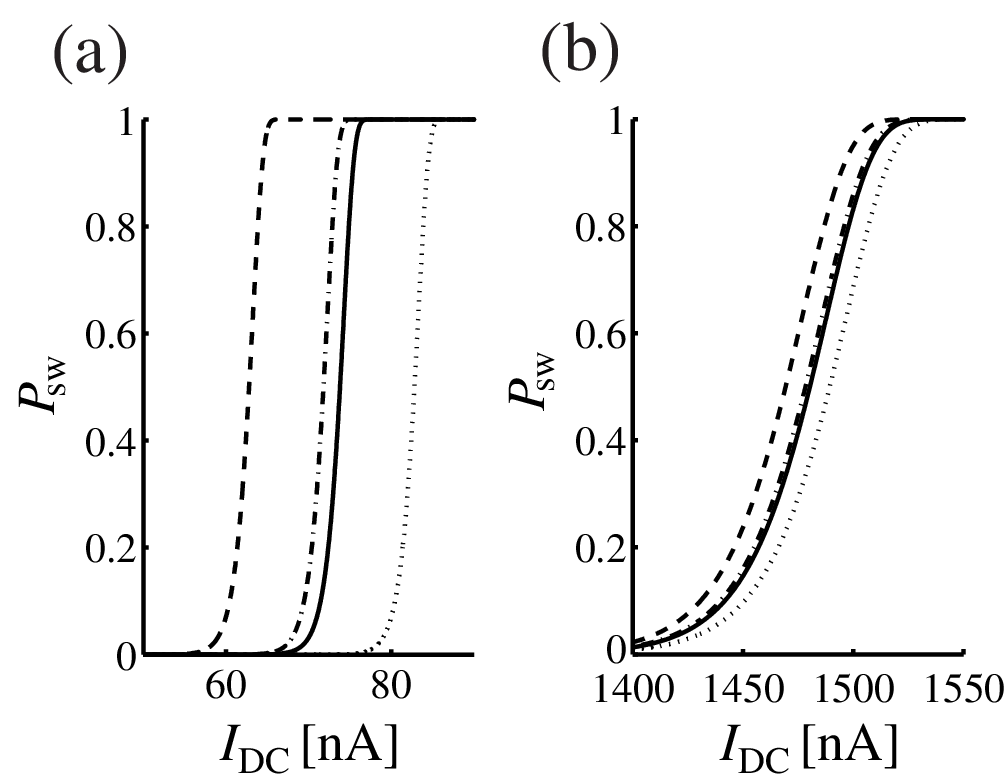}
\caption{\label{fig5}The probability of the system of Fig.~\ref{fig4} to switch to the normal conducting state as a function of
current~$I_\textrm{DC}$ with $10$~nA backward pumping (dashed line), $1$~nA backward pumping (dash-dotted line), $1$~nA forward pumping (solid
line), and $10$~nA forward pumping (dotted line). The parameters of the third junction are chosen to be (a) $I_{\textrm{c}3}=0.1$~$\mu$A and
$C_{\textrm{J}3}=200$~fF, and (b) $I_{\textrm{c}3}=2$~$\mu$A and $C_{\textrm{J}3}=15$~fF. The duration of the DC current pulse is $\tau=100$~$\mu$s.}
\end{figure}

We estimate the switching probability in the zero temperature limit as 
\be\label{sprob} P_\textrm{sw}=1-e^{-\tau\Gamma_\textrm{MQT}}.\ee The transition rate for the macroscopic quantum tunneling in a weakly
dissipative environment is given by
\be\label{gmqt}\Gamma_\textrm{MQT}=12\sqrt{6\pi}\frac{\omega_\textrm{p}}{2\pi}\sqrt{\du/(\hbar\omega_\textrm{p})}e^{-\frac{36\du}{5\hbar\omega_\textrm{p}}}
,\ee where $\du\approx\frac{2}{3}\ejc[2(1-\tilde{I}/I_{\textrm{c}3})]^{3/2}$ and
$\omega_\textrm{p}\approx\sqrt{8\ejc\ecc[2(1-\tilde{I}/I_{\textrm{c}3})]^{1/2}}/\hbar$~\cite{WeissBook} denote the barrier height and the plasma
frequency of the $3$rd junction, respectively. Figure~\ref{fig5} shows the switching probability given by Eq.~\eqref{sprob} as a function of the
DC current~$I_\textrm{DC}$ with forward and backward pumping for pumped currents of $10$~nA and $1$~nA which were found to be experimentally
feasible in Sec.~\ref{sec4}. In Fig.~\ref{fig5}(a), we have assumed that $I_{\textrm{c}3}=0.1$~$\mu$A and $C_{\textrm{J}3}=200$~fF. Since the
sluice is also assumed to be optimized for high Josephson energy, the additional assumption of the parameters of the third junction implies that
for example in standard shadow evaporation combined with electron beam lithography, three-angle evaporation is to be employed to obtain a
different transparency of the third junction as compared with the junctions used in the sluice. An alternative to the three-angle evaporation is
to install an additional capacitor in parallel with the third junction, as realized in Refs.~\cite{Cottet2002,Vion2002}. Although these
techniques in the fabrication of the sample are very feasible, we have plotted in Fig.~\ref{fig5}(b) the switching probability without them,
i.e., for $I_{\textrm{c}3}=2$~$\mu$A and $C_{\textrm{J}3}=15$~fF. To detect the pumping for the parameters of Fig.~\ref{fig5}(b), the relative
visibility in the variation of the applied DC current should be of the order of 0.1\% for $I_\textrm{p}=1$~nA and 1\% for $I_\textrm{p}=10$~nA.
We note that in our current experimental set up for other switching measurements than the proposed one, we have detected smaller than 0.1\%
shifts in the current histograms similar to the ones shown in Fig.~\ref{fig5}. Thus an accurate measurement of the pumped current for the
parameters of Fig.~\ref{fig5}(a) should be achievable.

The correction to the shift of the critical current due to finite residual Josephson energy~$E_\textrm{J}^\textrm{res}$ is of the order of
$\max(I_{\textrm{c}1},I_{\textrm{c}2})\epsilon$ which is small compared with the pumped current~$I_\textrm{p}$ for typical experimental
parameters. Thus we conclude that the proposed measurement scheme provides a convenient experimental arrangement for observation of Berry phase
in superconducting circuits.

Furthermore, we note that Fig.~\ref{fig4} strongly resembles the circuit used in the Saclay experiments on the artificial two-state atom
quantronium~\cite{Cottet2002,Vion2002}. The sluice in Fig.~\ref{fig4} corresponds to the quantronium and the other part of the circuit is
introduced for measuring the supercurrent through it. Since the supercurrent depends on the state of the quantronium, the switching of the
superconducting system into normal state can be employed to observe the state of the quantronium. Thus the supercurrent cannot be neglected as
we did in the derivation of Eq.~\eqref{idcx}, and the exact calculation of the switching becomes more difficult than in our case. For further
details, see Refs.~\cite{Ankerhold2003,Ithier2005}.

\section{Conclusions}\label{sec6}
In this paper, we analyzed a phase biased Cooper pair sluice, for which the pumped charge is closely related to the Berry phase accumulated in
one pumping cycle. We showed that the adiabaticity requirement of the sluice does not restrict the pumped current to be undetectably small if
the sluice is optimized according to the results derived. In fact, pumped currents of tens of nanoamperes might be achievable. Moreover, a lower
bound for the maximum pumped current allowed by the adiabaticity criterion for relative pumping error of $10^{-7}$ and standard experimental
parameters turned out to be of the order of $10$~pA. In Ref.~\cite{Niskanen2003}, a pumped current of~$100$~pA was obtained with the error rate
$10^{-7}$ using a dynamical approach, which suggests that the lower bound for the maximum pumped current obtained in this paper can be increased
by a more detailed derivation. Thus the sluice is a potential candidate for a metrological current standard.

We presented a detailed scenario for observing the pumped current, and showed that the visibility of the current in the proposed measurement is
expected to be high. In future work, we aim to experimentally observe the Berry phase employing the scheme given in this paper. Although we have
neglected sources of error, e.g., decoherence in our analysis, we expect that they are not dominant in the experimental
realization~\footnote{For a study of the effect of decoherence on the Berry phase, see Ref.~\cite{Whitney2005}, and references therein.}. In
fact, our goal is only to combine two already realized techniques: the Cooper pair pumping~\cite{Niskanen2005} and the switching measurement
with a large Josephson junction as an ammeter. Thus the scheme for observing the Berry phase in superconducting circuits presented here is
considered to be very promising.

\begin{acknowledgments}
We thank Academy of Finland for financial support. MM acknowledges the Finnish Cultural Foundation,Vilho, Yrj\"o, and Kalle V\"ais\"al\"a
Foundation, and FWJH acknowledges support from Institut Universitaire de France. We would like to express our appreciation to J.~Peltonen for
helpful comments.
\end{acknowledgments}

\bibliographystyle{prsty}
\bibliography{manu}

\end{document}